\documentclass[aps,prb,superscriptaddress,twocolumn,10pt,longbibliography]{revtex4-2}

\usepackage{amsmath}
\usepackage{graphicx}
\usepackage{bm}
\usepackage{color}

\date{\today}

\definecolor{darkblue}{rgb}{0.1,0.2,0.6}
\definecolor{darkred}{rgb}{0.8,0.1,0.2}
\usepackage[colorlinks,citecolor=darkblue,linkcolor=darkred,urlcolor=darkblue]
{hyperref}
\usepackage[all]{hypcap}

\usepackage{etoolbox}
\usepackage{mathtools}
\usepackage{amsmath}%
\usepackage{amsfonts}%
\usepackage{amssymb}
\usepackage[normalem]{ulem}

\begin{document}

\title{Many-body localization in a tilted potential in two dimensions}
\author{Elmer~V.~H.~Doggen}
\email[Corresponding author: ]{elmer.doggen@kit.edu}
\affiliation{\mbox{Institute for Quantum Materials and Technologies, Karlsruhe Institute of Technology, 76021 Karlsruhe, Germany}}
\affiliation{\mbox{Institut f\"ur Theorie der Kondensierten Materie, Karlsruhe Institute of Technology, 76128 Karlsruhe, Germany}}
\author{Igor~V.~Gornyi}
\affiliation{\mbox{Institute for Quantum Materials and Technologies, Karlsruhe Institute of Technology, 76021 Karlsruhe, Germany}}
\affiliation{\mbox{Institut f\"ur Theorie der Kondensierten Materie, Karlsruhe Institute of Technology, 76128 Karlsruhe, Germany}}
\affiliation{Ioffe Institute, 194021 St.~Petersburg, Russia}
\author{Dmitry~G.~Polyakov}
\affiliation{\mbox{Institute for Quantum Materials and Technologies, Karlsruhe Institute of Technology, 76021 Karlsruhe, Germany}}

\begin{abstract}
Thermalization in many-body systems can be inhibited by the application of a linearly increasing potential, which is known as Stark many-body localization. Here we investigate the fate of this phenomenon on a two-dimensional disorder-free lattice with up to $24 \times 6$ sites. Similar to the one-dimensional case, ``density-polarized'' regions can act as bottlenecks for transport and thermalization on laboratory timescales. However, compared to the one-dimensional case, a substantially stronger potential gradient is needed to prevent thermalization when an extra spatial dimension is involved. The origin of this difference and implications for experiments are discussed. We argue that delocalization is generally favored for typical states in two-dimensional Stark many-body systems, although nonergodicity can still be observed for a specific choice of initial states, such as those probed in experiments.
\end{abstract}

\maketitle

\section{Introduction}
Many systems encountered in nature obey the ergodic hypothesis, that is, each microstate consistent with fixed macroscopic thermodynamic variables according to the appropriate statistical ensemble is as likely as the other. However, some systems are nonergodic. Understanding the origin of nonergodicity in quantum systems \cite{DAlessio2016a} is of particular relevance to describing decoherence and the crossover from quantum to classical behavior. A paradigmatic example of nonergodicity in quantum many-body systems is many-body localization (MBL), which occurs at nonzero density of excitations, driven by the interplay between interactions and disorder \cite{Gornyi2005a, Basko2006a, Nandkishore2015a, Altman2015a, Abanin2017a, Alet2018a}.

Recently, interest has increased in studying many-body systems that exhibit nonergodicity even without the presence of any disorder \cite{Papic2015a,Smith2017a}. One such system consists of interacting particles under the influence of a linear potential, i.e., the interacting many-body analog of Wannier-Stark localization. Experimental realizations in one dimension have shown that localization can also persist in this case \cite{Scherg2021a,Guo2020a,Morong2021a}.

On the other hand, it has been observed experimentally \cite{Guardado2020a,Scherg2021a,Guo2020a,Morong2021a} that ``Stark-MBL'' systems can exhibit features of ergodic systems. This suggests a transition in such systems from a delocalized phase to a localized one at a certain critical value of the gradient of the linear potential. Numerical studies in one dimension have indicated that such a transition does indeed occur \cite{Schulz2019a,vanNieuwenburg2019a}, similarly to the predicted transition in disorder-driven MBL systems. However, as we have previously shown \cite{Doggen2021b}, within the purported delocalized region some states are, in fact, anomalously long-lived (see also Refs.~\cite{Yao2021a, Zisling2021a}).

\begin{figure*}[ht!]
    \centering
    \includegraphics[width=.67\columnwidth]{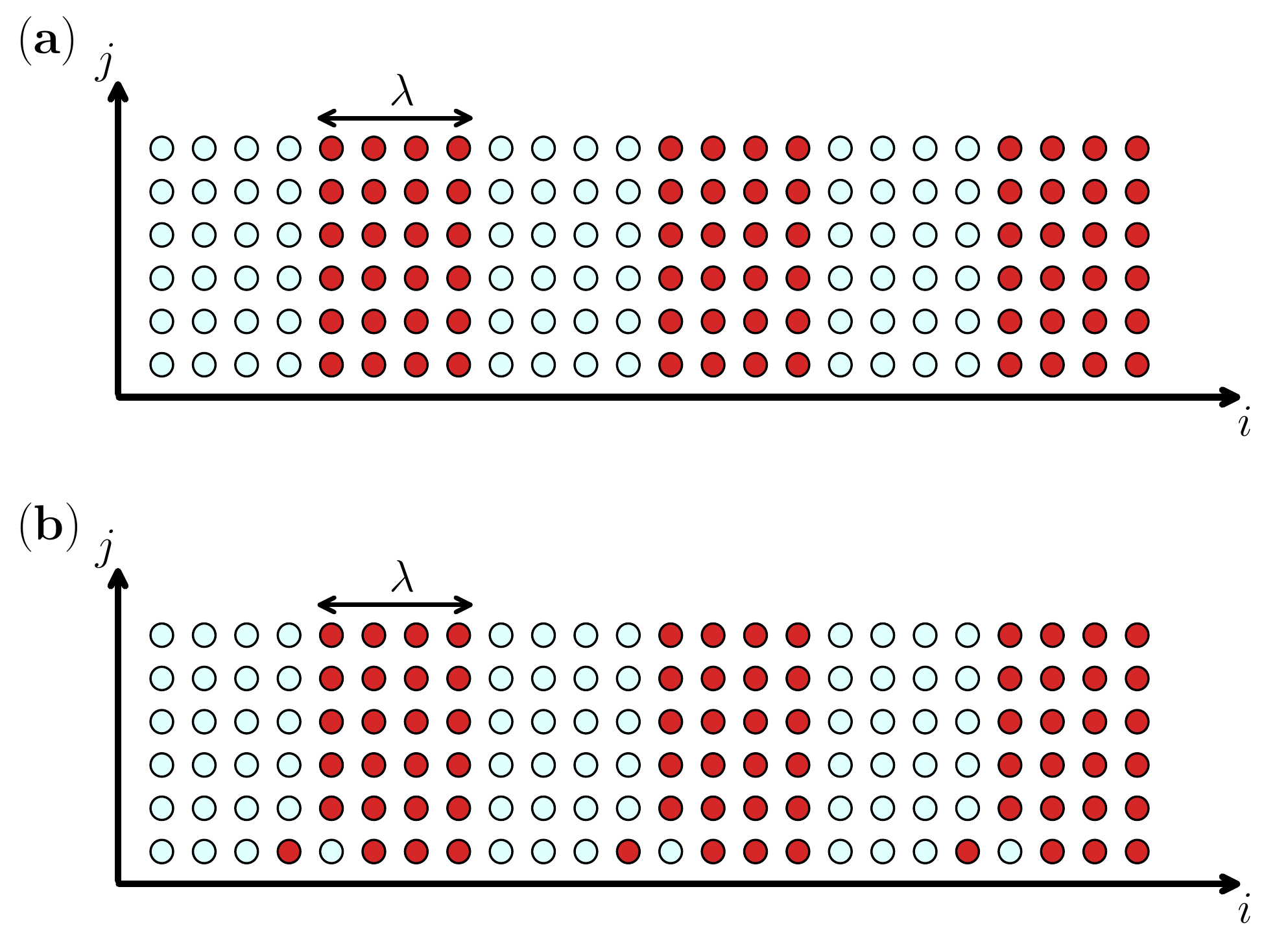}
    \includegraphics[width=.6\columnwidth]{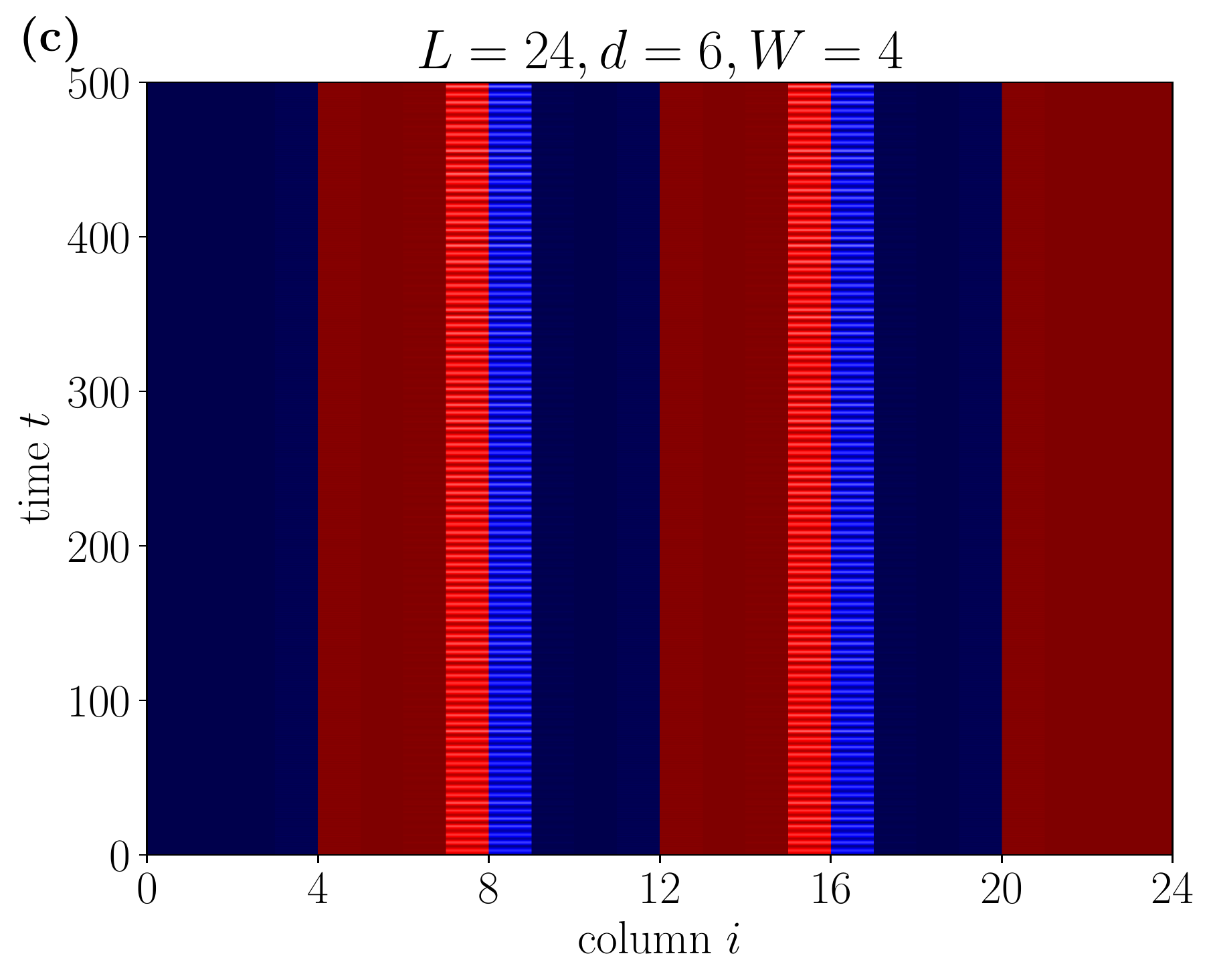}
    \includegraphics[width=.7\columnwidth]{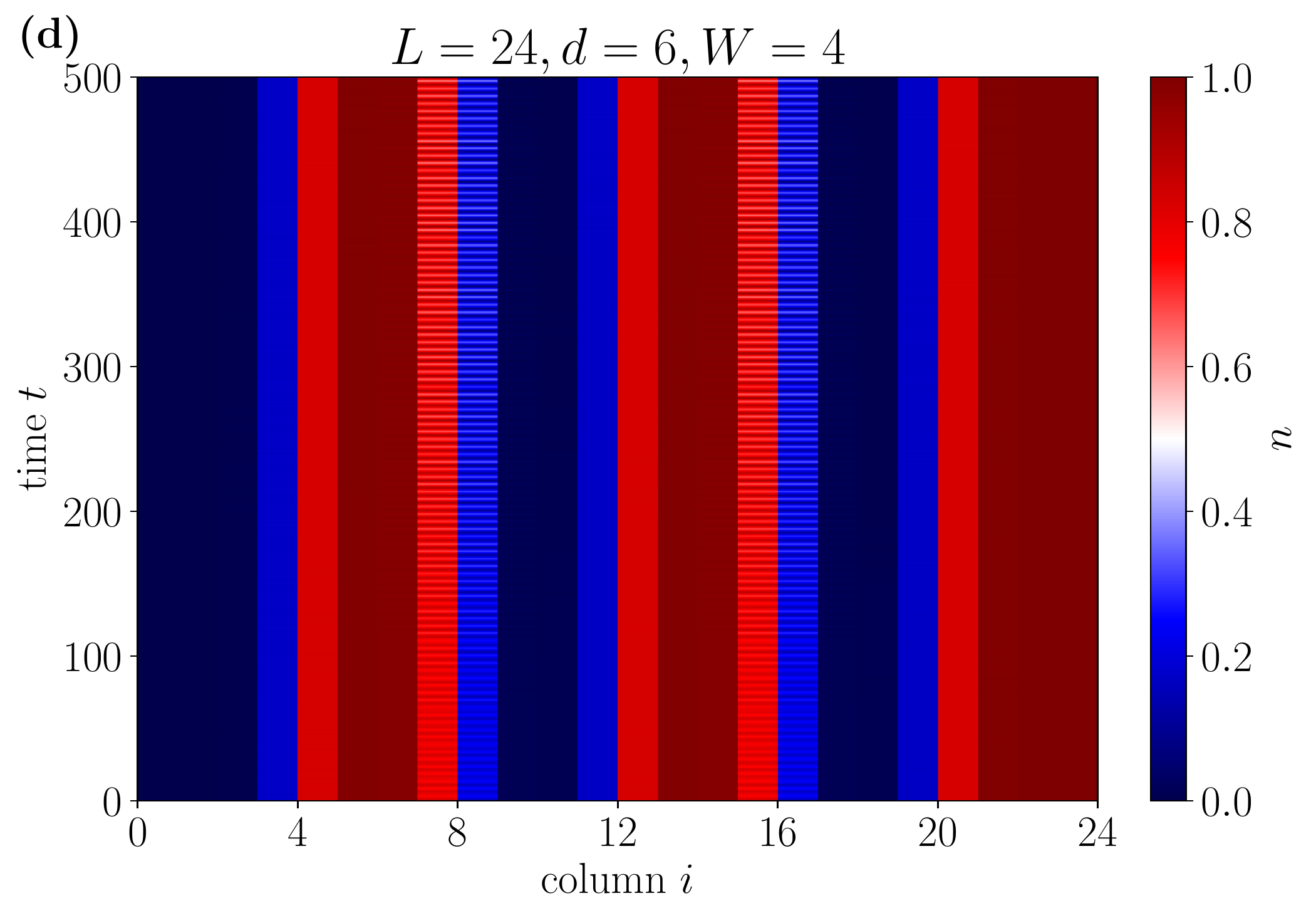}
    \caption{\textbf{(a,b)}: Schematics of the geometry and initial conditions. \textbf{(a)}: Charge-density wave (CDW) initial condition with wavelength $2\lambda$, where sites are either occupied (red dots) or unoccupied (light blue dots). The initial state is periodic in the $i$-direction, in the same direction as the potential gradient. \textbf{(b)}: As in panel \textbf{(a)}, but with a perturbation that breaks the translational invariance in the $j$-direction. \textbf{(c)}: Particle density $n$, averaged over sites within a given column (fixed $i$), as a function of time. The initial state is a unidirectional CDW as depicted in panel \textbf{(a)}, with $\lambda = 4$. \textbf{(d)}: As in panel \textbf{(c)}, but starting from the perturbed CDW initial state depicted in panel \textbf{(b)}. }
    \label{fig:diag}
\end{figure*}

One may understand this phenomenon through the Hilbert-space shattering (fragmentation) \cite{Sala2020a,Khemani2020a} that occurs in the limit of an infinitely large gradient of the potential, permitting a mapping to fractonic or constrained systems. Hilbert-space shattering implies that the Hilbert space of the system is divided into an exponentially large \cite{Khemani2020a} (in the system size) number of disconnected sectors, preventing thermalization. At a finite value of the potential gradient these sectors are connected, but in a sufficiently weak manner such that thermalization can be strongly suppressed on laboratory timescales. Importantly, in the one-dimensional (1D) case, the probability for the transport-blocking regions of an arbitrarily large length $\lambda$ to occur is unity in the thermodynamic limit, giving rise to a finite density of such regions. The thermodynamic limit is established in systems whose length is exponentially large in $\lambda$. 

A question of interest is to what degree nonergodicity in the various systems that exhibit it shares a similar origin and features. In this context, it is worthwhile to investigate the influence of geometry and dimensionality on localization-related phenomena. Disorder-driven MBL is only expected to be stable in one dimension, according to the avalanche theory of the MBL transition \cite{DeRoeck2017a,Thiery2017a,Morningstar2020a, Doggen2020a}. Within this theory, a vital role is played by rare weakly disordered regions, which lead to the emergence of growing ``ergodic spots'' that can eventually thermalize the whole system. Such regions are far more likely to occur in dimensions higher than one, destroying, in particular, MBL in two-dimensional (2D) systems in the thermodynamic limit. However, in the case of Stark MBL there are no possible rare configurations of the potential. From this perspective, Stark-MBL systems are more prone to localization, and the above distinction between 1D and 2D geometries is less prominent. 

On the other hand, despite the differences behind the physics of localization, some features familiar from disorder-driven MBL systems have been reported also in Stark-MBL systems, such as logarithmic growth of entanglement and Poissonian energy level statistics in the localized phase \cite{Schulz2019a, vanNieuwenburg2019a}. Furthermore, as we show below, a distinction between one- and higher-dimensional systems, as in disorder-driven MBL, is also present in the case of Stark MBL. This distinction is present because the suppression of transport in one dimension is related to blocking (polarized) regions, which occur with unit probability in the limit of large system sizes \cite{Doggen2021b}. However, this probability vanishes in the thermodynamic limit for 2D systems. It is the goal of this paper to sort out the similarities and differences between Stark MBL and conventional MBL with regard to the role of dimensionality.

From a technical point of view, a major obstacle is that the numerical complexity of a generic, unconstrained many-body quantum system on a lattice scales as $f^N$, where $N$ is the number of sites on the lattice and $f$ the number of local degrees of freedom. Exact algorithms can only handle system sizes up to $N \approx 25$, even in the simplest case $f = 2$ (e.g., a spin-$1/2$ system). This means that, in order to access meaningfully large systems, approximate methods need to be used. Recently, one such method---the time-dependent variational principle (TDVP) \cite{Haegeman2016a}---has proven to be exceptionally powerful, yielding reliable results for (almost) localized systems~\cite{Kloss2018a, Doggen2018a, Doggen2019a, Chanda2020a, Doggen2020a, Strkalj2021a, Doggen2021a, Doggen2021b}. This is true even up to relatively large times and system sizes, comparable to those of the experiment.
Here, we use the TDVP to elucidate the physics of Stark MBL in two dimensions.

\section{Model and Method}
We consider hard-core bosons on a 2D square lattice as described by the Hamiltonian:
\begin{equation}
 \mathcal{H} =  \sum_{\langle ij;i'j' \rangle} \left[ -\frac{J}{2} \left(b_{ij}^\dagger b_{i'j'} + \mathrm{H.c.}\right)  + U\hat{n}_{ij} \hat{n}_{i'j'} \right] + \sum_{ij} \epsilon_i \hat{n}_{ij},
  \label{eq:ham}
 \end{equation}
where $b_{ij}$ ($b^{\dagger}_{ij}$) is the annihilation (creation) operator for a boson on the site with indices $i \in [1,L],\,j \in [1,d]$ and $\hat{n}_{ij} = b_{ij}^\dagger b_{ij}$. The summation over $\langle ij;i'j'\rangle$ is restricted to nearest neighbors, with open (periodic) boundary conditions in the $i$ ($j$)-direction. In the following, we use units with $\hbar = 1$ 
and choose $J = 1$ for the energy and time scales. Moreover, we set the interaction strength $U = 1$. The on-site potential varies only in the $i$-direction, namely $\epsilon_i = Wi$, where $W$ is the potential gradient. This model is similar to the one studied in a recent experiment \cite{Guardado2020a}, except that we consider the simpler case of hard-core bosons instead of two-component fermions.
  
Dynamics governed by Eq.~(\ref{eq:ham}) is computed, using the TDVP, up to time $t = 500$. The TDVP belongs to the matrix-product-state (MPS) class of algorithms \cite{Schollwock2011a}, a type of variational representation of the many-body wave function with a controllable error.  The TDVP dynamics obeys the Schr\"odinger-like equation:
\begin{equation}
\frac{d}{dt} |\psi \rangle = -i \mathcal{P}_\mathrm{MPS}\mathcal{H}|\psi\rangle, \label{eq:tdvp}  
\end{equation}
where $\mathcal{P}_\mathrm{MPS}$ projects the dynamics onto the variational manifold. The number of independent parameters in the manifold scales with the bond dimension $\chi$, which is the main control parameter used to verify convergence of the algorithm. A major appeal of this method, compared to other MPS algorithms, is that globally conserved quantities are conserved in the numerical procedure, enhancing accuracy. We employ the hybrid one-site and two-site implementation of the TDVP~\cite{Doggen2020a, Doggen2021b} in a parallelized fashion (details and benchmarks are presented in Appendix).

\section{Results}
With the TDVP, we compute dynamics starting from an initial product state, using a sufficiently large bond dimension. We consider two different initial states $|\psi\rangle$, which are product states in the particle occupation basis. The first, depicted schematically in Fig.~\ref{fig:diag}a, is a steplike charge-density wave (CDW) configuration with wavelength $2\lambda$ and dimensions $L$ in the $i$-direction and $d$ in the $j$-direction. This choice corresponds to the one employed in the experimental realization of Ref.~\cite{Guardado2020a}. We furthermore consider an additional initial state, depicted in Fig.~\ref{fig:diag}b, where the aforementioned state is perturbed, breaking the translational invariance in the $j$-direction. We approach the two-dimensional limit by considering a width of up to $6$ lattice sites.

Choosing the potential to be translationally invariant perpendicular to the axis of the CDW is appealing from the perspective of investigating localization properties, because it is expected that this setup is the most amenable to delocalization. Furthermore, with this arrangement, we can directly investigate the fate of the long-lived 1D states studied in Ref.~\cite{Doggen2021b} upon increasing the width of the system, thus going towards the 2D case.

Aside from considering the expectation values of particle densities $n_{ij} \equiv \langle b_{ij}^\dagger b_{ij} \rangle$, we consider the memory of the initial state, as quantified using the imbalance $\mathcal{I}$, an experimentally accessible quantity \cite{Schreiber2015a}:
\begin{equation}
    \mathcal{I}(t) = \frac{4}{Ld} \sum_{ij} \Big[n_{ij}(t) - 1/2 \Big] \Big[ n_{ij}(t=0) - 1/2 \Big]. \label{eq:imba}
\end{equation}
A state that is unchanged from the initial state obeys $\mathcal{I} = 1$, while for a delocalized state at half filling $\mathcal{I} = 0$.

We further consider the bipartite von Neumann entropy of entanglement $S$ \cite{Laflorencie2016a}:
\begin{align}
S(t) = \max_{A}[-\mathrm{Tr}(\rho_A \ln \rho_A)], \nonumber \\
\rho_A \equiv \mathrm{Tr}_B |\psi (t) \rangle \langle \psi (t) |. \label{eq:entropy}
\end{align}
Here $\mathrm{Tr}_B$ traces out the degrees of freedom corresponding to part $B$. We choose the bipartition between subsystems $A$ and $B$ such that the entropy is the maximum (this is the meaning of $\max_{A}$ above) of all the possible bipartitions, which turns out to correspond to the position of a domain wall in the setups in Figs.~\ref{fig:diag}a and b (e.g., for the parameters in Fig.~\ref{fig:diag}c, the maximum of $S$ is achieved at $i = 8$ and $16$).

Let us first consider dynamics for a particular choice of parameters $L = 24$, $d = 6$, $\lambda = 4$, and $W = 4$, as shown in Fig.~\ref{fig:diag}a. After a brief initial evolution (see Appendix), dynamics is frozen up to $t = 500$ hopping times, without any appreciable change in the state, apart from regular oscillations, see Fig.~\ref{fig:diag}c. The possibility of transverse dynamics, therefore, does not appear to lead to delocalization in the $i$-direction.

\begin{figure}
    \centering
    \includegraphics[width=\columnwidth]{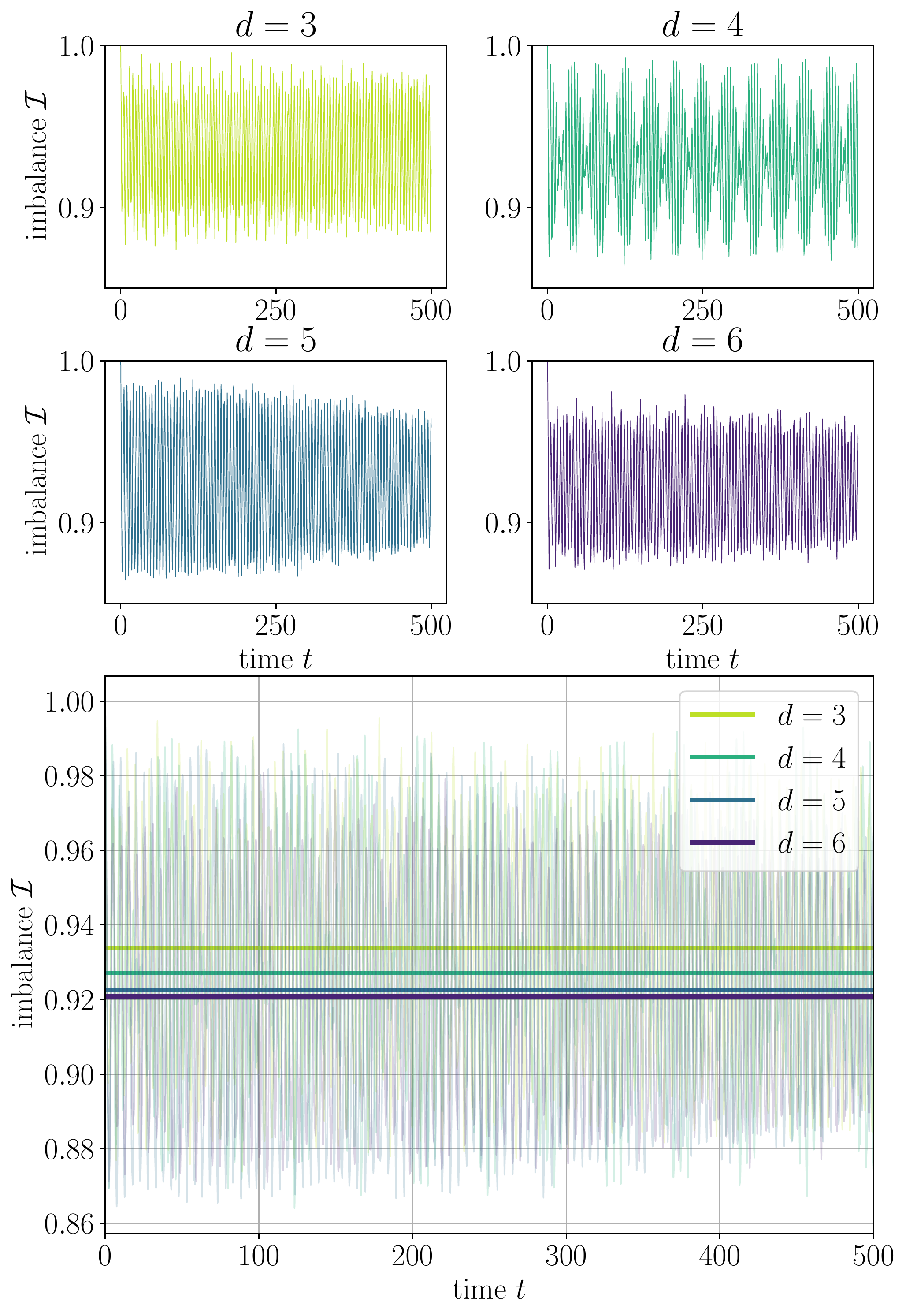}
    \caption{Imbalance \eqref{eq:imba} as a function of time for various widths of the system, $d \in [3,6]$. Other parameters are the same as in Fig.~\ref{fig:diag}. Top panels show individual imbalance curves for $d \in [3,6]$. The thick lines in the bottom panel show the leading behavior of the imbalance without oscillations, using a low-pass filter. }
    \label{fig:imba}
\end{figure}

In Fig.~\ref{fig:imba}, we show the imbalance as a function of time. Because the initial density pattern mostly survives (see Fig.~\ref{fig:diag}c), $\mathcal{I}$ remains close to 1. Strong oscillations in time are present, which are more pronounced for smaller widths $d$. This is due to the dynamics being constrained to only a few sites. The average value (solid lines in Fig.~\ref{fig:imba}, obtained through a Fourier transform, analogous to Ref.~\cite{Yao2021a}) is only weakly dependent on system size, suggesting convergence to an asymptotic value for larger system sizes. This implies that localization survives up to long times, similar to the 1D case.

We now consider the von Neumann entropy of entanglement \eqref{eq:entropy}. Similar to the imbalance studied above, the entropy shows an initial rapid change and then saturates at a fixed value. This saturation value of the entropy scales with the system width $d$, roughly as $S \propto d$. Such behavior is expected, because the dynamics is constrained only in the $i$-direction. Indeed, the number of involved sites scales linearly with $d$, which leads to the same scaling for the entropy.

\begin{figure}
    \centering
    \includegraphics[width=\columnwidth]{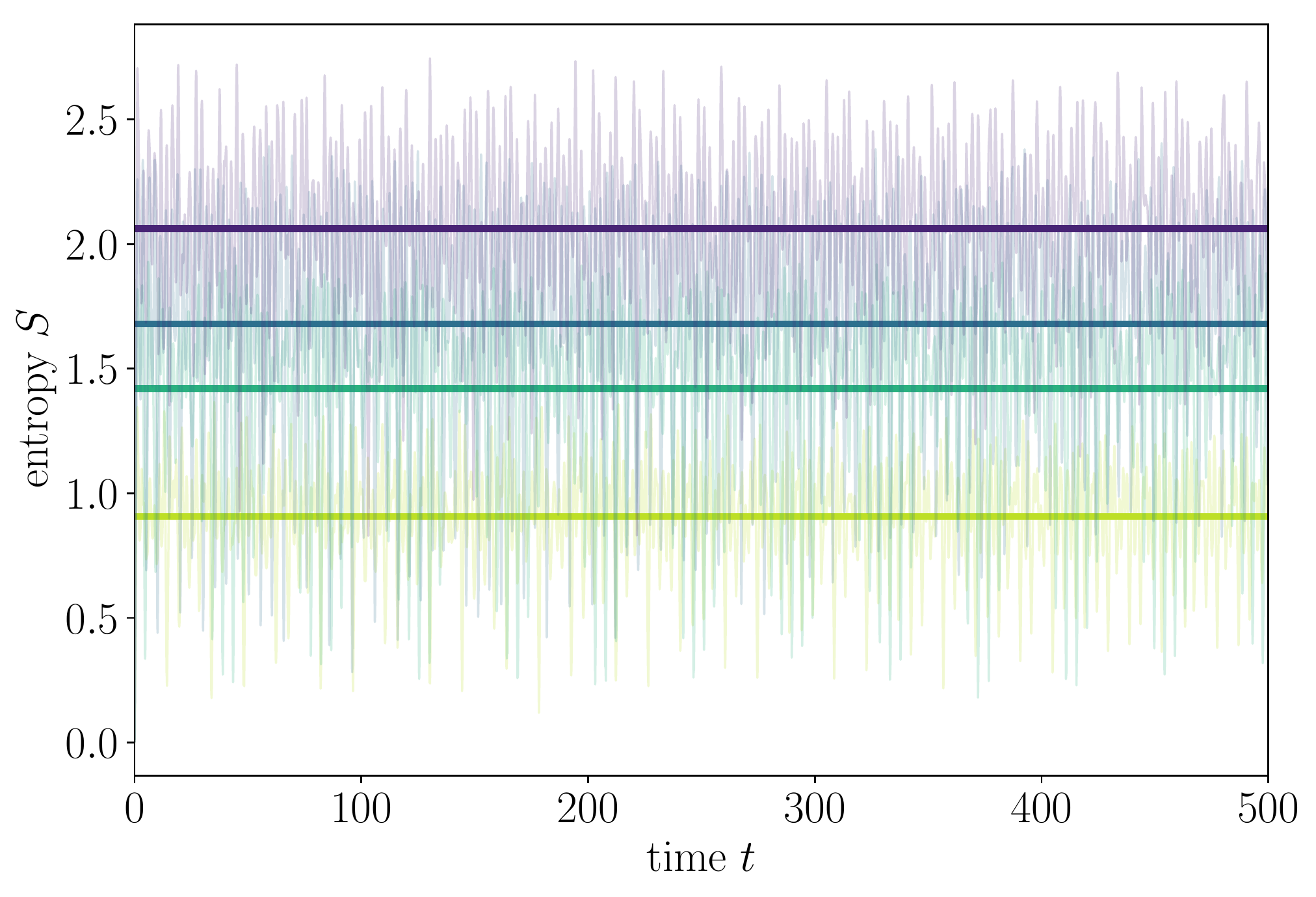}
    \caption{Entanglement entropy \eqref{eq:entropy} as a function of time, for the same parameters as in Fig.~\ref{fig:imba}.}
    \label{fig:entropy}
\end{figure}

Perturbing the initial state in the manner depicted in Fig.~\ref{fig:diag}b breaks translational invariance in the $j$-direction and introduces additional broadening of the left domain walls (at $i=4,\, 12$, and $20$; we choose the same parameters $L$, $d$, $\lambda$, and $W$ as for Fig.~\ref{fig:diag}a). Comparing the two cases (Figs.~\ref{fig:diag}c and d), we observe highly similar dynamics at the right (unperturbed) domain walls, whereas the dynamics is different for the left (perturbed) boundaries. Nonetheless, the dynamics in Fig.~\ref{fig:diag}d appears frozen over the observed timescales, with no sign of macroscopic thermalization.
Notably, in both Figs.~\ref{fig:diag}c and d, the spatial range of the domain-wall melting in the $i$-direction does not exceed the width of the system $d=6$. In other words, any 2D subblocks of size $6\times 6$
(see Figs.~\ref{fig:diag}a and b) can be considered as non-thermalized on the timescale of observation.

\begin{figure}
    \centering
    \includegraphics[width=\columnwidth]{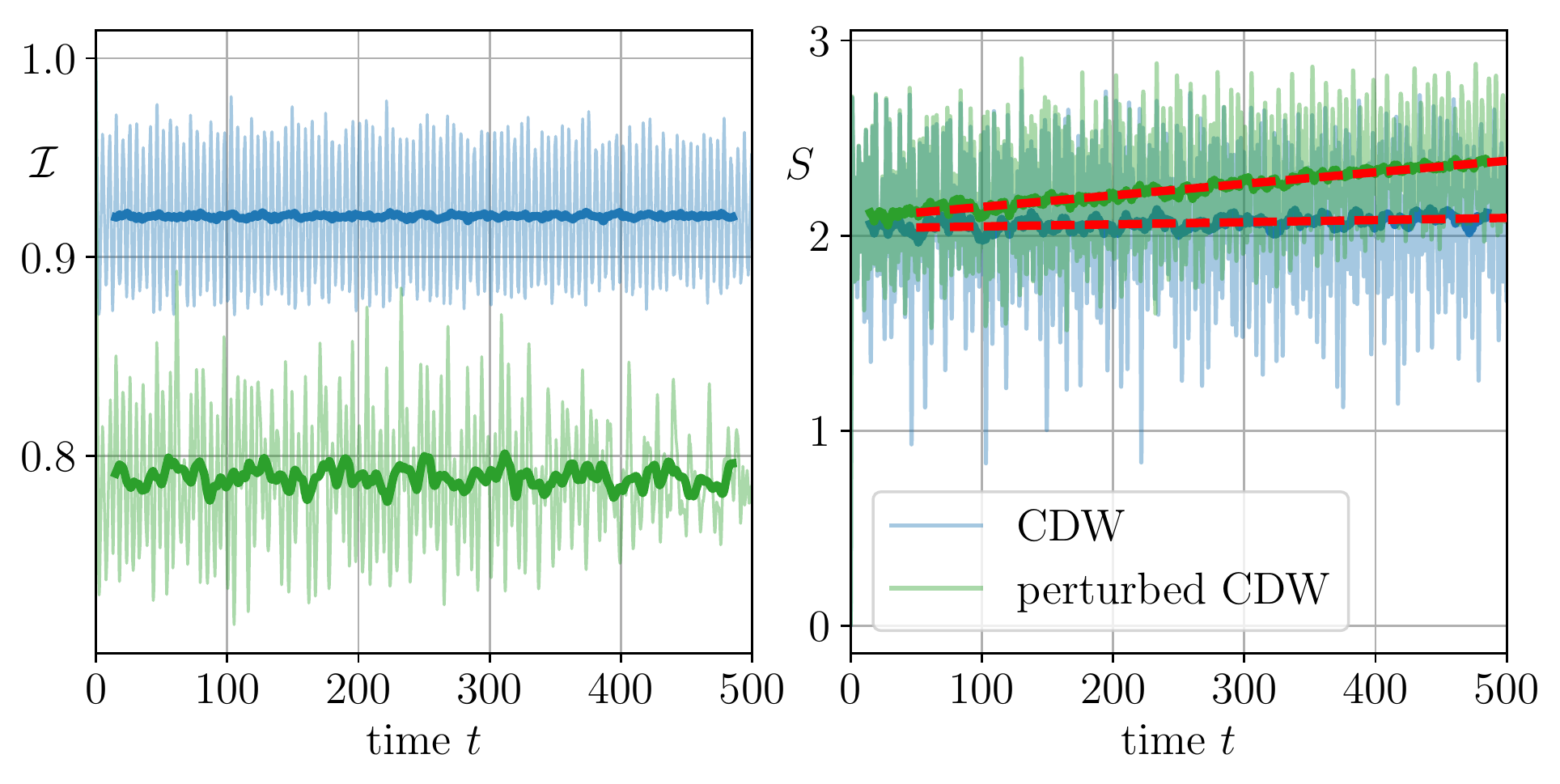}
    \caption{Comparison of the imbalance (left panel) and entropy (right panel) dynamics for the same choice of parameters, $L = 24$, $d = 6$, $W = 4$, $\lambda = 4$, and $\chi = 384$, but different choices of initial condition: the charge density wave (Fig.~\ref{fig:diag}a) and perturbed charge density wave (Fig.~\ref{fig:diag}b). The dashed red lines indicate a linear fit to the curve $S(t) = at + b$, yielding the values $a_\mathrm{CDW} = (11 \pm 5) \cdot 10^{-5}$ and $a_\mathrm{pCDW} = (59 \pm 4) \cdot 10^{-5}$ for the CDW and perturbed CDW respectively (95\% confidence intervals). The thick lines represent filtered data using a Savitzky-Golay polynomial fitting procedure of third order \cite{Savitzky1964a}.}
    \label{fig:comparison}
\end{figure}

Let us now inspect the dynamics of the imbalance and entropy in the perturbed and unperturbed cases in more detail. In Fig.~\ref{fig:comparison}, a comparison between the CDW initial condition and the perturbed CDW is shown. We note two essential differences. Firstly, in the perturbed case, the imbalance $\mathcal{I}$ initially drops to a lower value, $1-\mathcal{I} \sim 1/d$. After the drop, similarly to the unperturbed case, no significant decay of the imbalance is observed. 
Secondly, the time dependence of the entropy is markedly different. In the unperturbed case, the entropy quickly reaches a plateau at $S\approx 2$ from the initial value $S=0$, after which there is a barely noticeable increase in time. In the perturbed case, a substantially faster increase is visible at the second stage. The maximum value of the entropy is reached at the domain walls with the $|10\rangle$ configuration found at column indices $i = 8$ and $16$ (right domain walls), while the perturbation is at the left domain walls. Hence, correlations due to the perturbation do penetrate through the domain wall. The behavior is reasonably well fitted by a linear dependence, depicted in Fig.~\ref{fig:comparison}, which provides a better fit than a logarithmic dependence over the depicted time window. Despite this steady growth of entanglement, no significant decay of the imbalance is observed, suggesting long-lived stability of such localized states. Such behavior is not contradictory: in the case of disordered MBL in one dimension, entropy growth is thought to be logarithmic, while transport remains frozen \cite{Bardarson2012a}.

\begin{figure}
    \centering
    \includegraphics[width=\columnwidth]{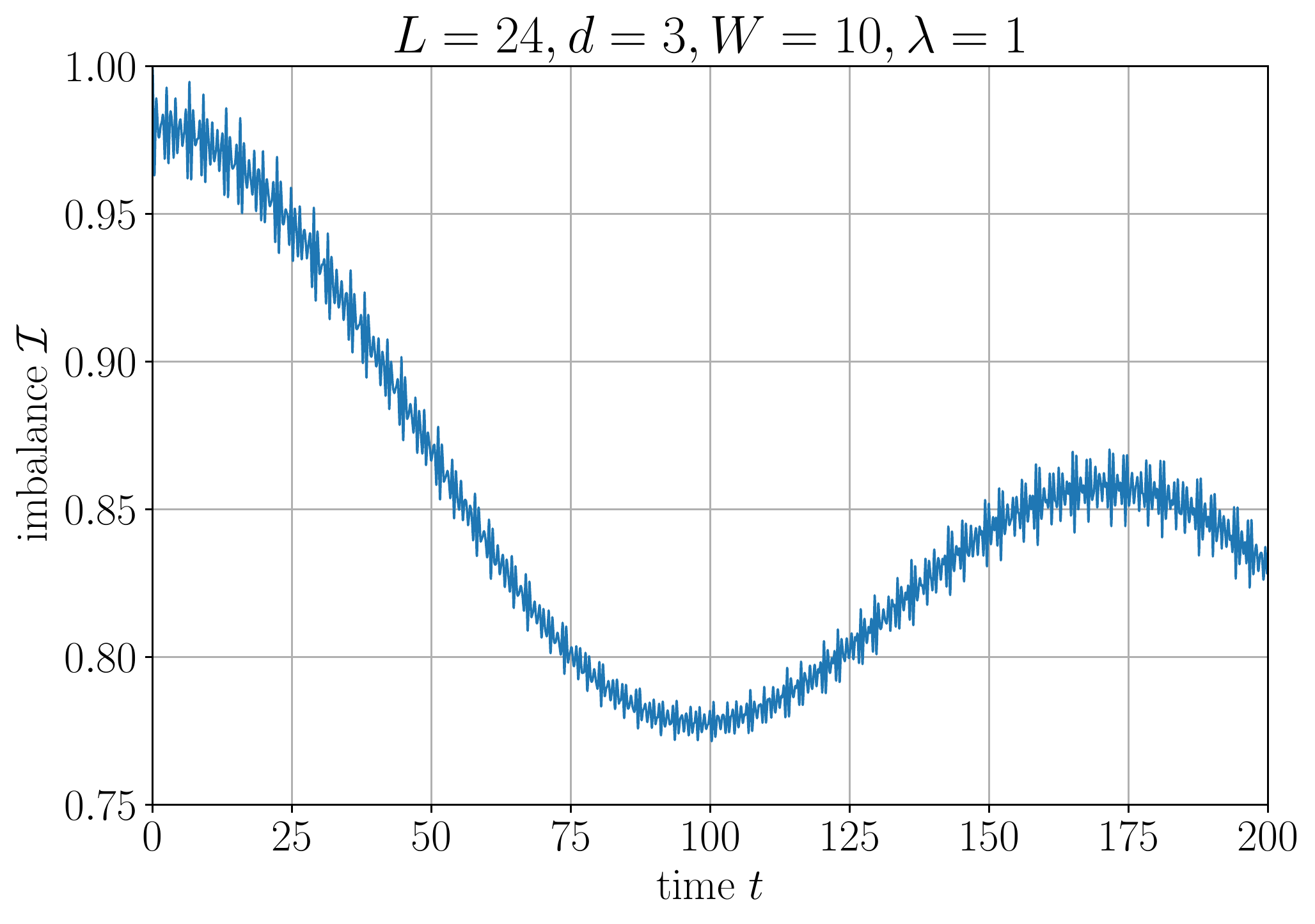}
    \caption{Imbalance dynamics in the case of a short-wavelength charge-density wave, with $\lambda = 1$. Note the choice of a stronger gradient $W = 10$.}
    \label{fig:lam1}
\end{figure}

If the period of the CDW is reduced, the polarized striped regions are less effective at inhibiting transport. In Fig.~\ref{fig:lam1}, we show the case where $\lambda = 1$ (and no perturbation to the CDW), with $L = 24$ and $d = 3$. Even though the width of the system is limited, we can observe a dramatic quantitative difference compared to dynamics for the 1D case~\cite{Schulz2019a, vanNieuwenburg2019a,Doggen2021b}. Namely, while in the 1D case, already a modest value $W \gtrsim 1$ is sufficient to observe clear saturation on these timescales, we do not observe such saturation in the quasi-1D case even at a much stronger value of the tilt, $W = 10$.

This difference is understandable as follows. In the 1D case, violations of the eigenstate thermalization hypothesis~\cite{DAlessio2016a} result from specific nonergodic states identified in Ref.~\cite{Doggen2021b}. These states with a blocking region of length exceeding a given $\lambda$
(which is a decreasing function of the field $W$) have measure zero in the whole Hilbert space in the limit $L \rightarrow \infty$, but still occur with a unit probability in the subspace of random product states. More specifically, they occur with a spatial density that scales as $2^{-\lambda}$.

For a quasi-1D system with width $d$, however, these blocking regions require polarized regions of area $d\lambda$. The probability of finding such a state scales as $2^{-d\lambda}$. In the picture presented in Ref.~\cite{Doggen2021b}, this means that localization is less enduring with respect to tuning the potential gradient away from $W \rightarrow \infty$ (recall that in this limit there is a mapping to a constrained, nonergodic system~\cite{Khemani2020a}). In the fully 2D limit, $d=L \rightarrow \infty$, the appearance of a fully blocking region then becomes vanishingly unlikely, in stark contrast to the 1D case.

\section{Conclusion and outlook}
In one dimension, the application of a linearly increasing potential induces localization, which survives the introduction of interactions (Stark many-body localization). Here, we have shown that long-lived localized states exist also in higher dimensions, namely, on a 2D lattice. This behavior is in line with the notion of Hilbert-space shattering, following a mapping to a constrained system that becomes exact at infinitely large potential gradient.  

However, it is noteworthy that the values of the field $W$ required to observe localization in higher dimensions are substantially larger than in the 1D case \cite{Doggen2021b}, where we observed long-lived localization in the 1D analog of the model \eqref{eq:ham} even for $W = 0.3$. The additional dimension, therefore, aids thermalization of the system. We can explain this dependence on dimensionality by noting that the blocking polarized regions represent an exponentially smaller part of the whole Hilbert space as a function of width $d$, compared to the 1D case. Therefore, there is a parallel to the ``standard'' type of MBL in a disordered potential, in which dimensionality is argued to play a crucial role \cite{DeRoeck2017a, Morningstar2020a, Doggen2020a} in that it determines the importance of rare fluctuations of disorder responsible for delocalization. In the Stark-MBL case, no disorder is present, but ``rare events''---rare \emph{states} with local constraints---also play a crucial role, favoring, in contrast to the rare ergodic spots in the disordered case, localization.
(In this sense, they are similar to transport-hindering rare events of the Griffiths type~\cite{Nandkishore2015a, Altman2015a}.)
The relative number of such states as a fraction of the Hilbert space is greatly suppressed for higher dimensions, again in contrast to the disordered case, where the ergodic spots are more probable with increasing dimensionality. Remarkably, the proliferation of rare ergodic regions in conventional 2D MBL and the suppression of blocking regions in 2D Stark MBL both have a delocalizing effect. The key difference between 1D and 2D Stark MBL is that, in two dimensions, the polarized blocking regions are expected to be irrelevant for the thermalization of \emph{typical} (product) states, in contrast to the 1D case  \cite{Doggen2021b}.

At smaller values of the gradient $W$, an analytical description, assuming an incoherent (hydrodynamic) picture, predicts subdiffusive transport~\cite{Guardado2020a, Zhang2020b}. This leads to strong growth of the entanglement, and is, therefore, extremely demanding for the numerical simulations based on matrix product states. The experiment of Ref.~\cite{Guardado2020a} does find, however, an exponential decrease of the delocalization rate in the observed subdiffusive regime as a function of the size of polarized regions, suggesting a trend toward localization for the CDW initial states. It may nonetheless be difficult to observe robust localization in a 2D system, as it is more challenging to prepare a cleanly polarized plaquette as opposed to a 1D charge-density wave. Indeed, as we have seen, a single (hole) defect in any one site at the boundary of the plaquette region quickly destroys polarization in the direction perpendicular to the field gradient~(see Appendix).

An intriguing open question is whether the numerically observed localization represents a genuine long-lived localized phase, or a transient ``prethermal'' state \cite{Abanin2017b}. 
Contrary to the one-dimensional case \cite{Doggen2021b}, we observe growth of entanglement with time, while transport remains frozen. This is potentially a signature of such a prethermal regime at timescales far beyond the numerically accessible range. Note, however, in disordered MBL systems the growth of entanglement is characteristic of both the ergodic and nonergodic states \cite{Bardarson2012a}.
We stress that, in the context of this work, by ``MBL'' we mean long-lived localized states -- the weakest possible criterion for MBL. The stability criteria for the MBL phase and the properties of the transition in the thermodynamic limit are still debated even for 1D disordered MBL \cite{Panda2020a, Kiefer2020a, Luitz2020a,Kiefer2021,Ghosh2021, Sierant2019a, Suntajs2020a, Abanin2021a, Morningstar2020a, Sels2021a}. It is difficult to address questions pertaining to the thermodynamic limit through experimental or numerical means, especially in two dimensions, so that further analytical work in this direction is needed.

\section*{Acknowledgments}
We thank F.~Pollmann and P.~Sala for useful discussions. Simulations were performed using the TeNPy library \cite{tenpy}, version \texttt{0.7.2}.

\appendix

\section{Numerical details}  \label{sec:appendix}

In this Appendix, we discuss technical details of the numerical simulations, as well as provide several benchmarks and a comparison to a different algorithm. In this work, we have employed matrix product states (MPS) simulations, a class of variational algorithms for solving many-body problems \cite{Schollwock2011a}, in which the number of variational parameters is controlled by the bond dimension $\chi$.

\begin{figure}
    \centering
    \includegraphics[width=\columnwidth]{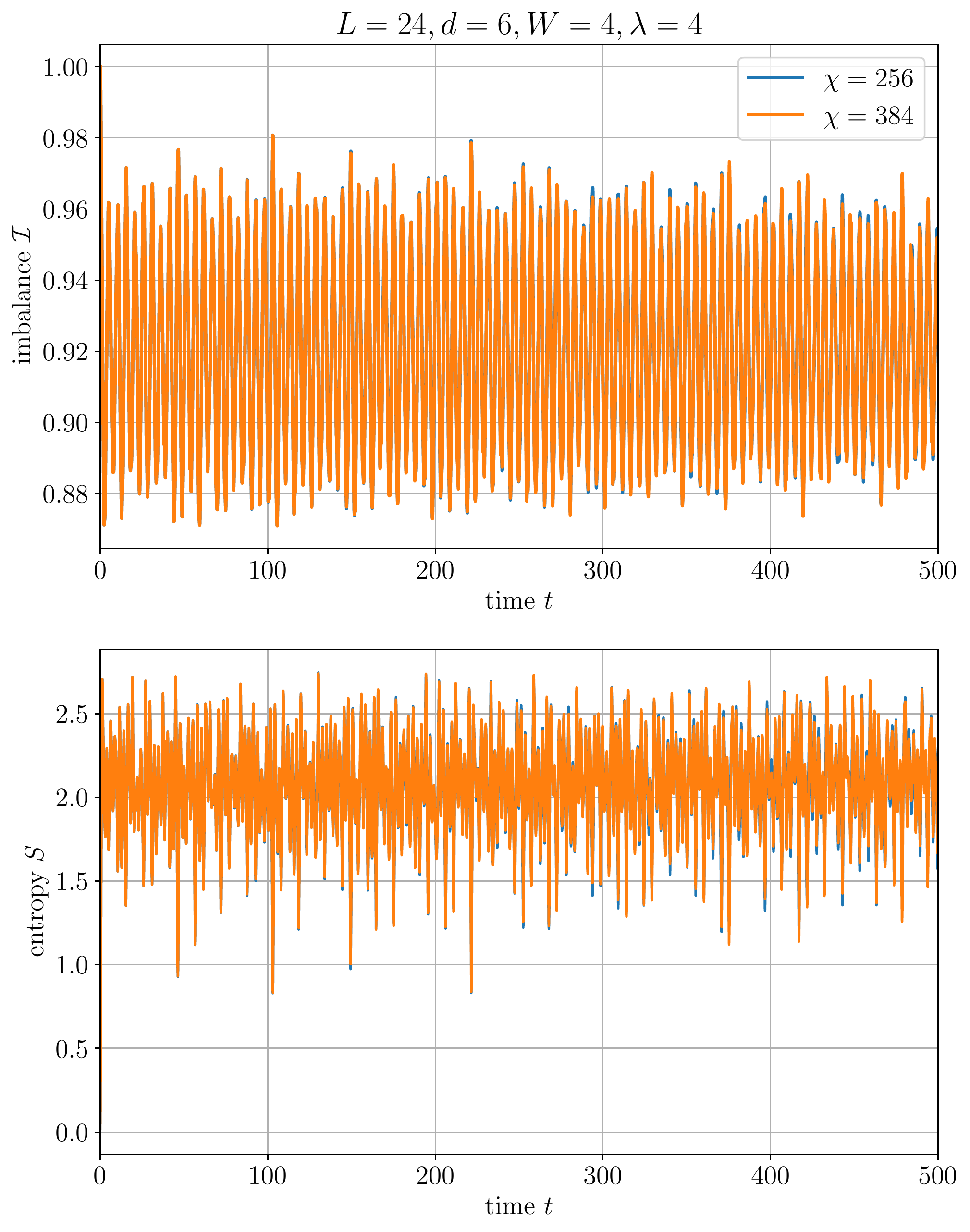}
    \caption{Comparison of the dynamics in the case of the CDW initial condition for different choices of the bond dimension $\chi = \{ 256, 384 \}$, for both the imbalance (top panel) and the entropy (bottom panel).}
    \label{fig:chibench}
\end{figure}

We employ the time-dependent variational principle (TDVP) \cite{Haegeman2016a} in a hybrid implementation (the same as used in Refs.~\cite{Doggen2020a,Doggen2021b}), where the two-site algorithm, which allows expansion of $\chi$ at every time step, is used up to $\tau = \max(2, \tau_\chi)$, where $\tau_\chi$ is the time in which $\chi$ has reached the maximum set value. After the time $t = \tau$, the remainder of the dynamics is computed using the single-site algorithm. This algorithm does not allow further expansion (or reduction) of the bond dimension, but has the benefit that the energy is globally conserved by the dynamics. This is opposed to other MPS-based algorithms, in which the various truncation procedures leads to violations of energy conservation, even for a time-independent Hamiltonian, which results in accumulating errors. Loosely speaking, one can identify this difference as the difference between implicit and explicit numerical integration schemes for solving partial differential equations. 
Instead of the parallel implementation used for disorder \cite{Doggen2020a}, here we employ parallelization of the Intel Math Kernel Library (MKL) routines for an additional speedup.

\subsection{Bond dimension}

Let us first consider the dependence of the result on the bond dimension $\chi$ of the MPS. Recall that $\chi$ controls the number of variational parameters in the MPS, allowing for stronger entanglement throughout the system as $\chi$ increases. We should then expect the result of the simulation to converge for sufficiently high $\chi$. A comparison for the choices $\chi = \{ 256, 384 \}$ is shown in Fig.~\ref{fig:chibench}. The results are in good agreement, with only a small discrepancy visible at late times.

\begin{figure}
    \centering
    \includegraphics[width=\columnwidth]{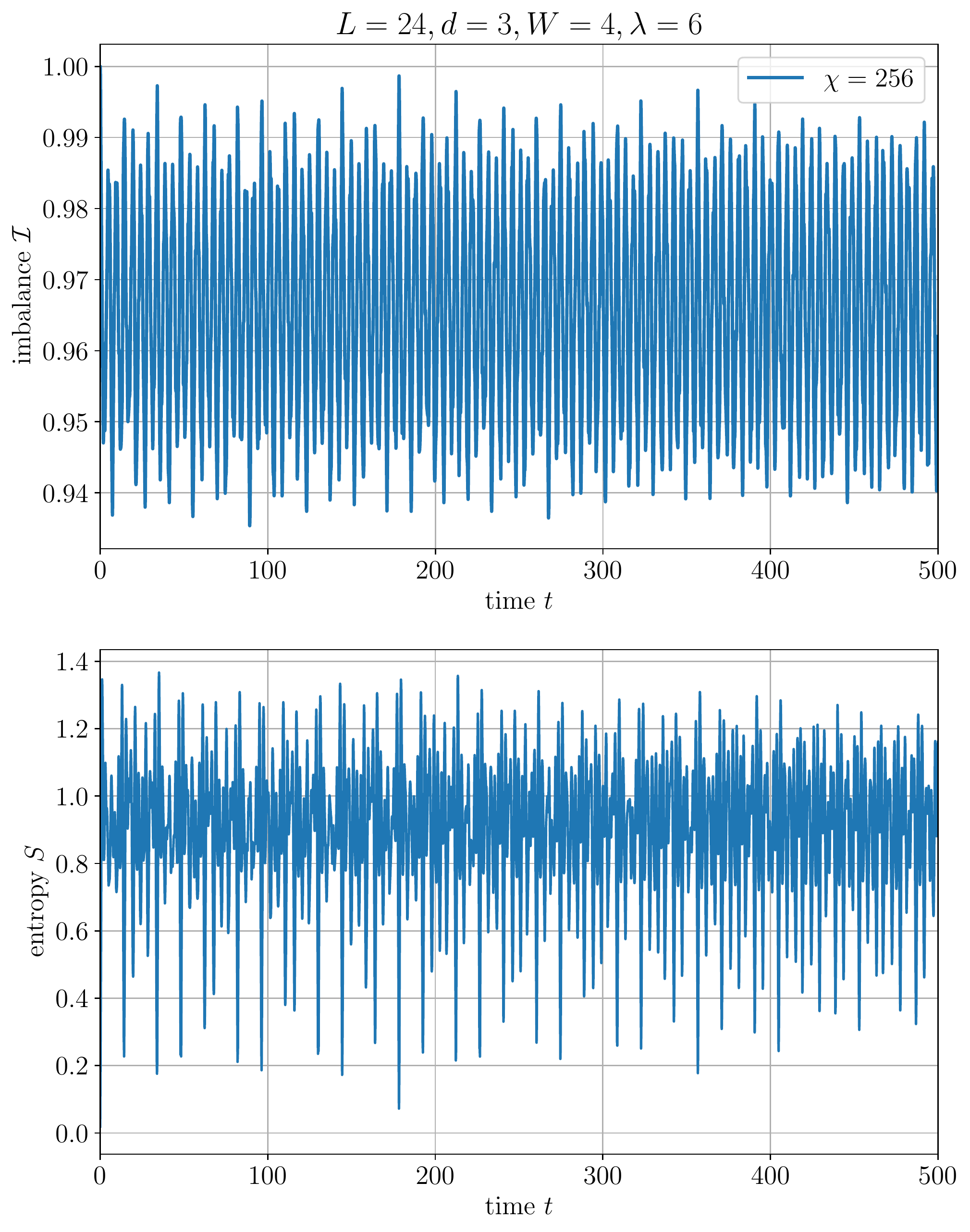}
    \caption{Imbalance dynamics (top) and entropy (bottom) for a choice of the initial condition corresponding to $\lambda = 6$.}
    \label{fig:diff_lambda}
\end{figure}

\subsection{Influence of the wavelength $\lambda$}

In the following, we consider the effect of changing the wavelength $\lambda$ of the initial charge-density wave. As long as we are in the localized phase, we should expect localization to remain robust upon increasing $\lambda$; this increases the length of polarized regions in the system. Results for the parameters $L = 24$, $d = 3$, $W = 4$, and $ \lambda = 6$ are shown in Fig.~\ref{fig:diff_lambda}. Comparison to the case $\lambda = 4$ in the main text indeed reveals the results are mostly unchanged: the maximum bipartite entropy is approximately the same. The imbalance also appears to saturate at a finite value, which in this case is slightly closer to 1, corresponding to the reduced number of domain wall borders.

\begin{figure}
    \centering
    \includegraphics[width=\columnwidth]{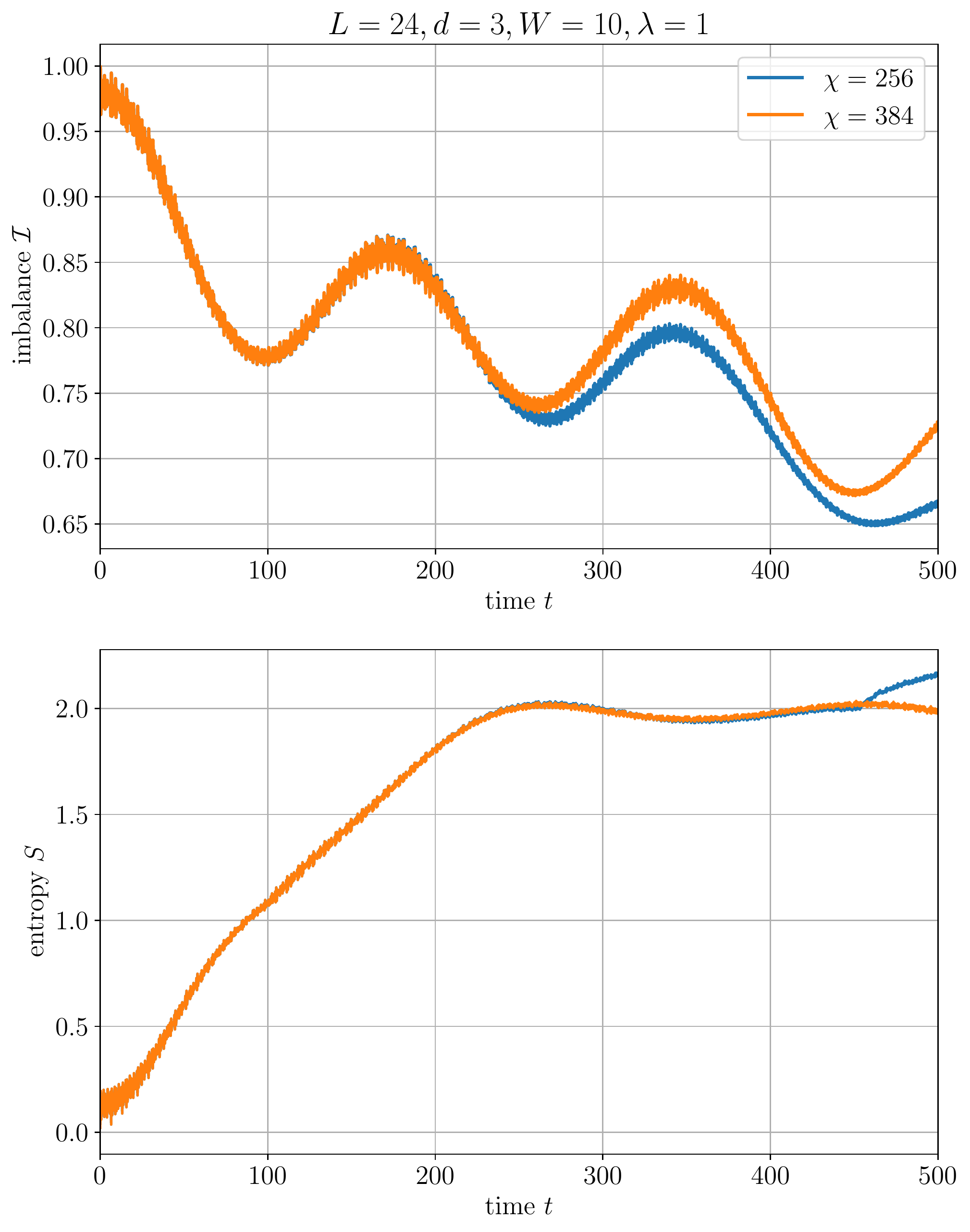}
    \caption{Imbalance dynamics (top) and entropy (bottom) for $\lambda = 1$, with a larger value of the tilt ($W = 10$) compared to the main text. Convergence with $\chi$ is lost at $t \approx 200$.}
    \label{fig:neel_init}
\end{figure}

Of interest is also the case where the wavelength is relatively small. If $\lambda = 1$, we obtain the so-called (columnar) N{\'e}el state as an initial condition, where columns are initially occupied and unoccupied in an alternating fashion. This state should be among the most susceptible to delocalization (out of the possible states with translational invariance in the $j$-direction), as is indeed observed numerically (see Fig.~\ref{fig:neel_init} and the main text). Even for a much larger choice of the tilt $W = 10$, the system tends to delocalize over time, and convergence with bond dimension is lost around $t \approx 200$ due to the growth of entanglement.

\subsection{Influence of system size $L$}

\begin{figure}
    \centering
    \includegraphics[width=\columnwidth]{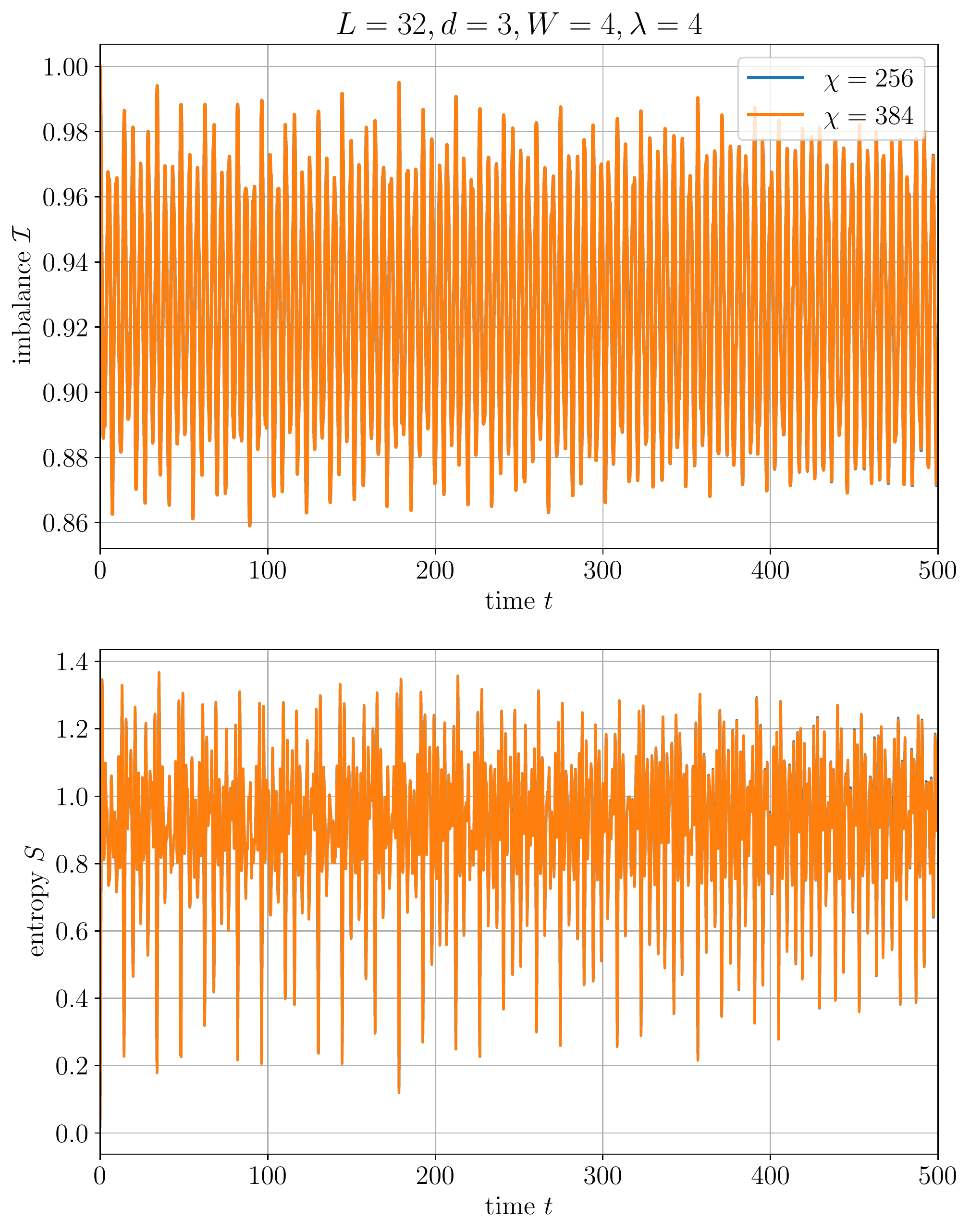}
    \caption{Imbalance dynamics (top) and entropy (bottom) for a different choice of system length compared to the main text, $L = 32$.}
    \label{fig:diff_length}
\end{figure}

Similarly to the above section, we can investigate the effect of changing system size. As per the same reasoning as for what happens in the case of increasing $\lambda$, little should change upon increasing the system size in the localized regime. The results for the choice $L = 32$, $d = 3$, $W = 4$, and $\lambda = 4$ are shown in Fig.~\ref{fig:diff_length}. Again, the results are highly similar to the choice $L = 24$ shown in the main text.

\subsection{Dynamics in the transverse direction}

\begin{figure}
    \centering
    \includegraphics[width=\columnwidth]{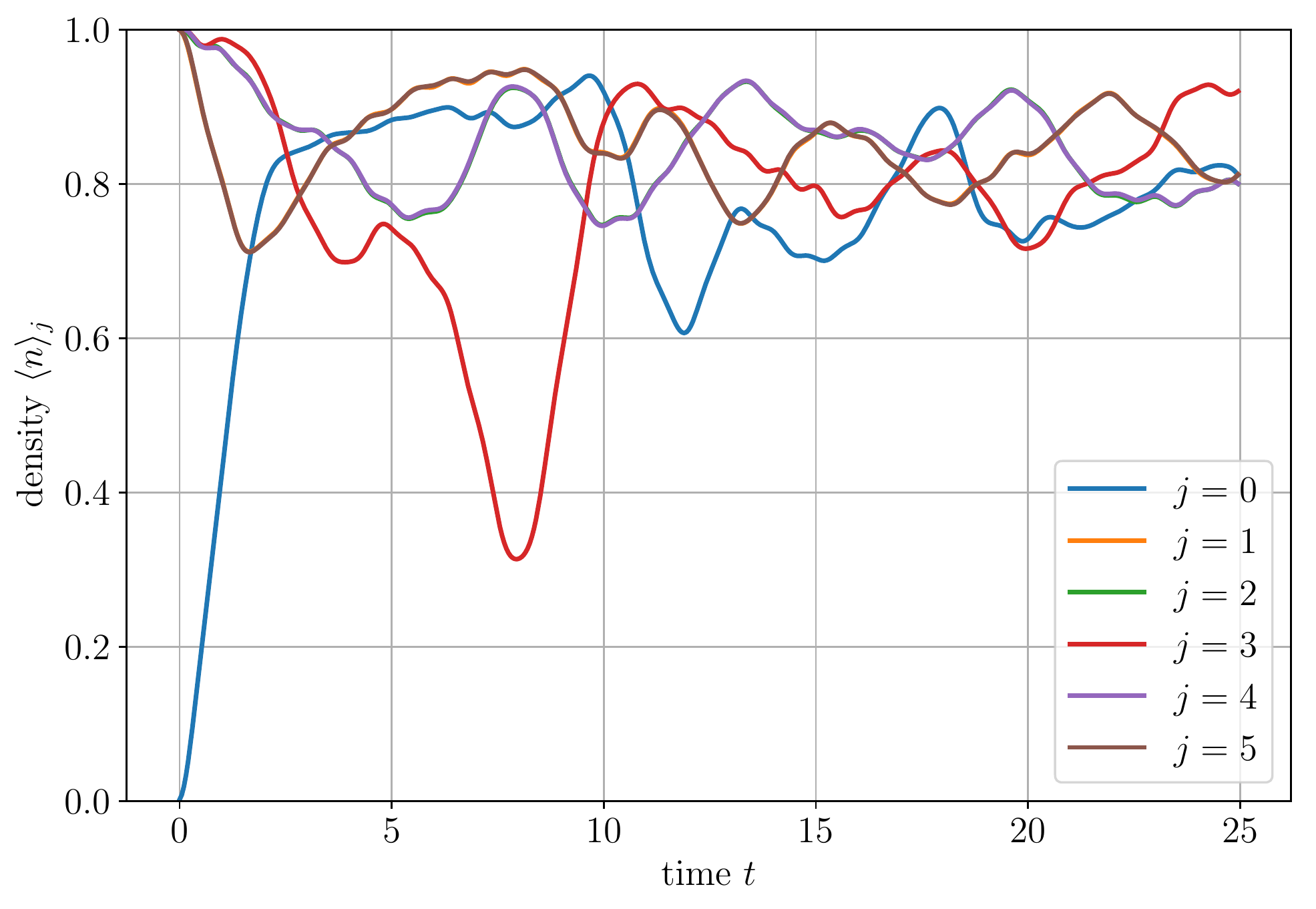}
    \caption{Dynamics for the on-site particle density $\langle n \rangle_j$ for a fixed row $i=5$ in the case of the perturbed CDW initial condition. Parameters are $L = 24$, $d = 6$, $W = 4$, $\lambda = 4$, and $\chi = 384$. Shown is the early time window $t \in [0, 25]$.}
    \label{fig:indiv1}
\end{figure}

\begin{figure}
    \centering
    \includegraphics[width=\columnwidth]{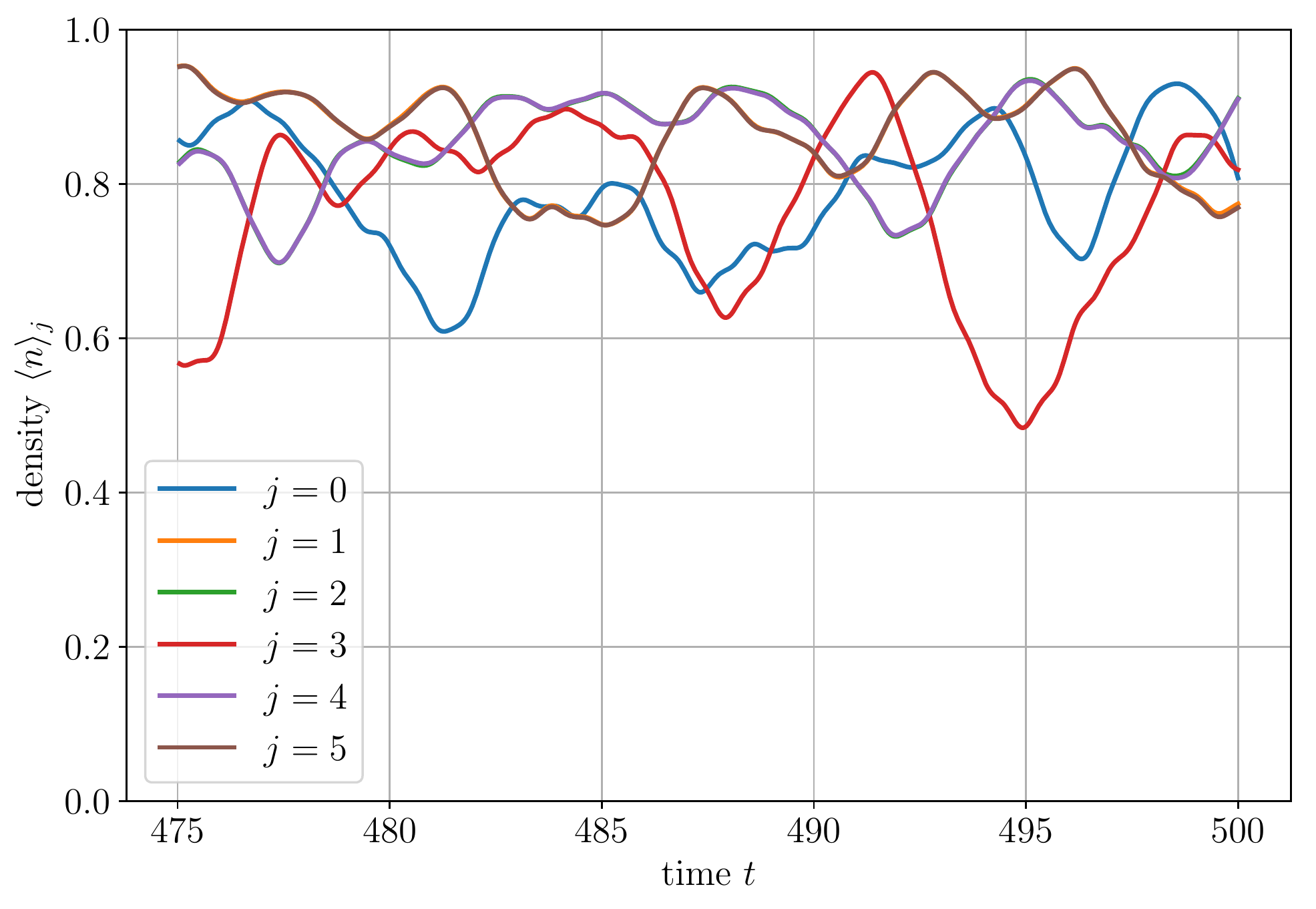}
    \caption{As Fig.~\ref{fig:indiv1}, but at a late time window $t \in [475, 500]$.}
    \label{fig:indiv2}
\end{figure}

It is instructive, also for the purposes of benchmarking, to investigate the dynamics of individual sites. Let us consider the case of the perturbed CDW as discussed in the main text. In the transverse direction, no potential gradient is present. Hence, particles are allowed to move freely in this direction, and we should expect rapid ``thermalization'' -- albeit restricted only to this transverse dimension. This is precisely what is found, as depicted in Figs.~\ref{fig:indiv1} and \ref{fig:indiv2}. On a relatively modest timescale $O(d)$ the site densities reach a value $1-\langle n \rangle \sim 1/d$, corresponding to the average density in the initial state. Around this average value, there are oscillations that are not damped because the dynamics is unitary and the system is closed.

Note that due to symmetry in the initial condition, the dynamics for the site pairs $j = \{1,5\}$ and $j = \{2, 4\}$ is identical. Since this symmetry is not explicitly imposed by the algorithm (in fact, the symmetry is broken through the mapping from a 2D lattice to a 1D chain, which breaks translational symmetries in the transverse direction), this provides another benchmark for the method. Only a very small difference in the densities is found, barely visible on the scale of the plot, even at the latest time window $t \in [475, 500]$ (Fig.~\ref{fig:indiv2}).

\subsection{Quadratic modulation of the potential}

In Refs.\cite{Schulz2019a, Taylor2020a} it has been argued that adding a small perturbation to the potential, taken to be in the form of a small quadratic term, can dramatically influence the localization properties of the model. The reason put forward is that the unperturbed linear potential permits resonant processes, enhancing delocalization. However, in Ref.~\cite{Yao2021b} the effect of this type of perturbation on the melting of domain walls has been considered, and the authors have not found crucial differences caused by the quadratic modulation. Here we show the effect of such a quadratic modulation in the 2D case, as considered in the main text. To wit, the full potential is now given by:
\begin{equation}
    \epsilon_{i, \mathrm{quad}} = Wi - \alpha i^2, 
\end{equation}
where $\alpha = 0.01$. Hence, the gradient of the field as a whole is approximately $W$ close to the left end of the system, but for $W = 4$ the potential reaches zero at $L = 20$, and then becomes negative. This is, thus, a rather significant perturbation of the potential.

\begin{figure}
    \centering
    \includegraphics[width=\columnwidth]{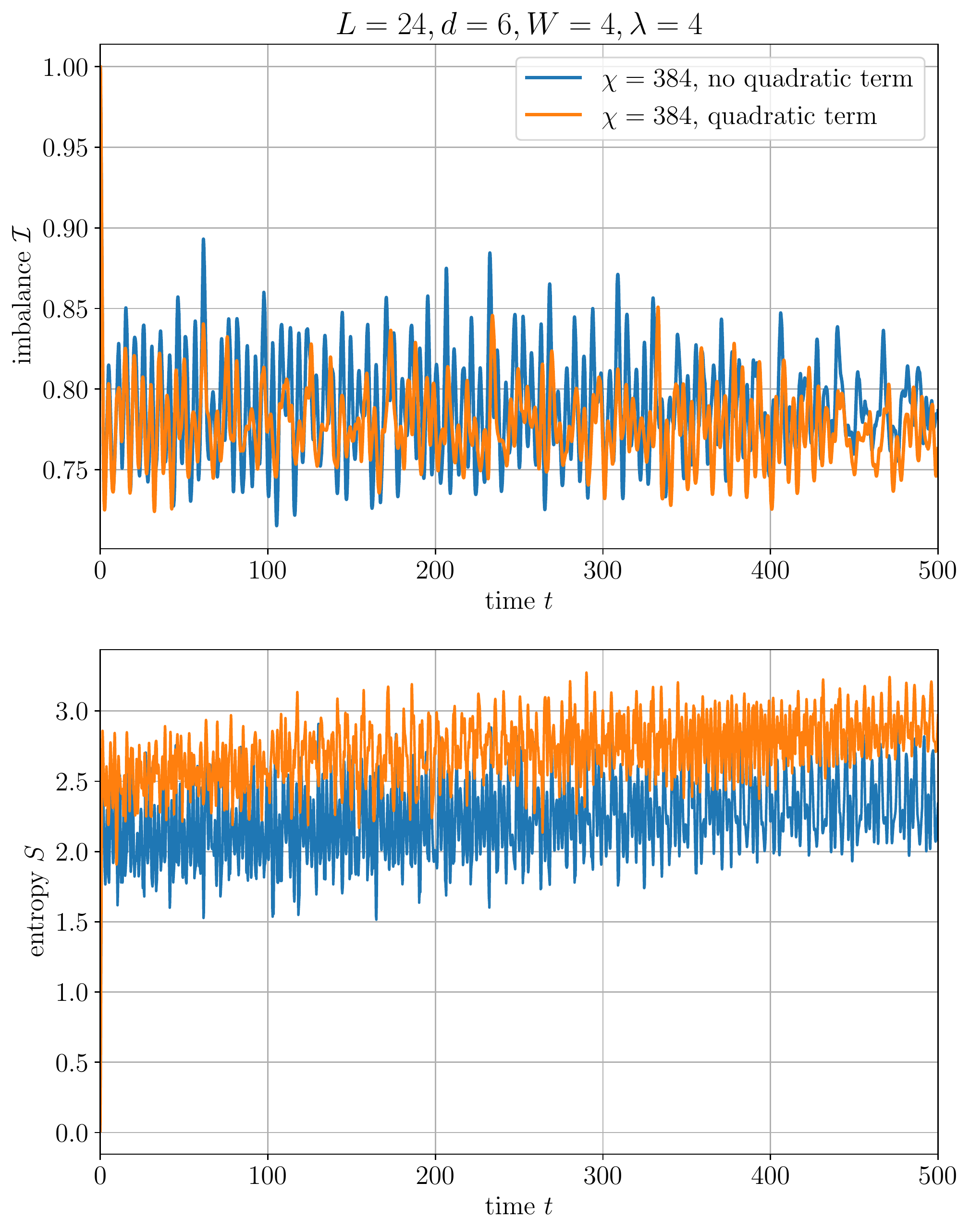}
    \caption{Comparison of dynamics in the case where a small quadratic perturbation is present, compared to the case where it is absent. The initial condition is chosen to be the perturbed CDW.}
    \label{fig:quad}
\end{figure}

The results for the dynamics in both cases $\alpha = 0$ and $\alpha = 0.01$ are shown in Fig.~\ref{fig:quad}. Here the initial condition with a perturbed CDW is chosen, corresponding to the green line in Fig.~\ref{fig:comparison} of the main text. The addition of the quadratic potential reduces the magnitude of oscillations in the imbalance, which can be attributed to the breaking of the aforementioned resonances. However, the qualitative behavior appears unchanged, with only a slightly lower imbalance and slightly higher entropy in the case of the quadratic modulation. The increased delocalization can be attributed to the effectively weaker gradient for the rightmost domain wall, enhancing the melting thereof \cite{Yao2021b}.

\begin{figure}[!htb]
    \centering
    \includegraphics[width=\columnwidth]{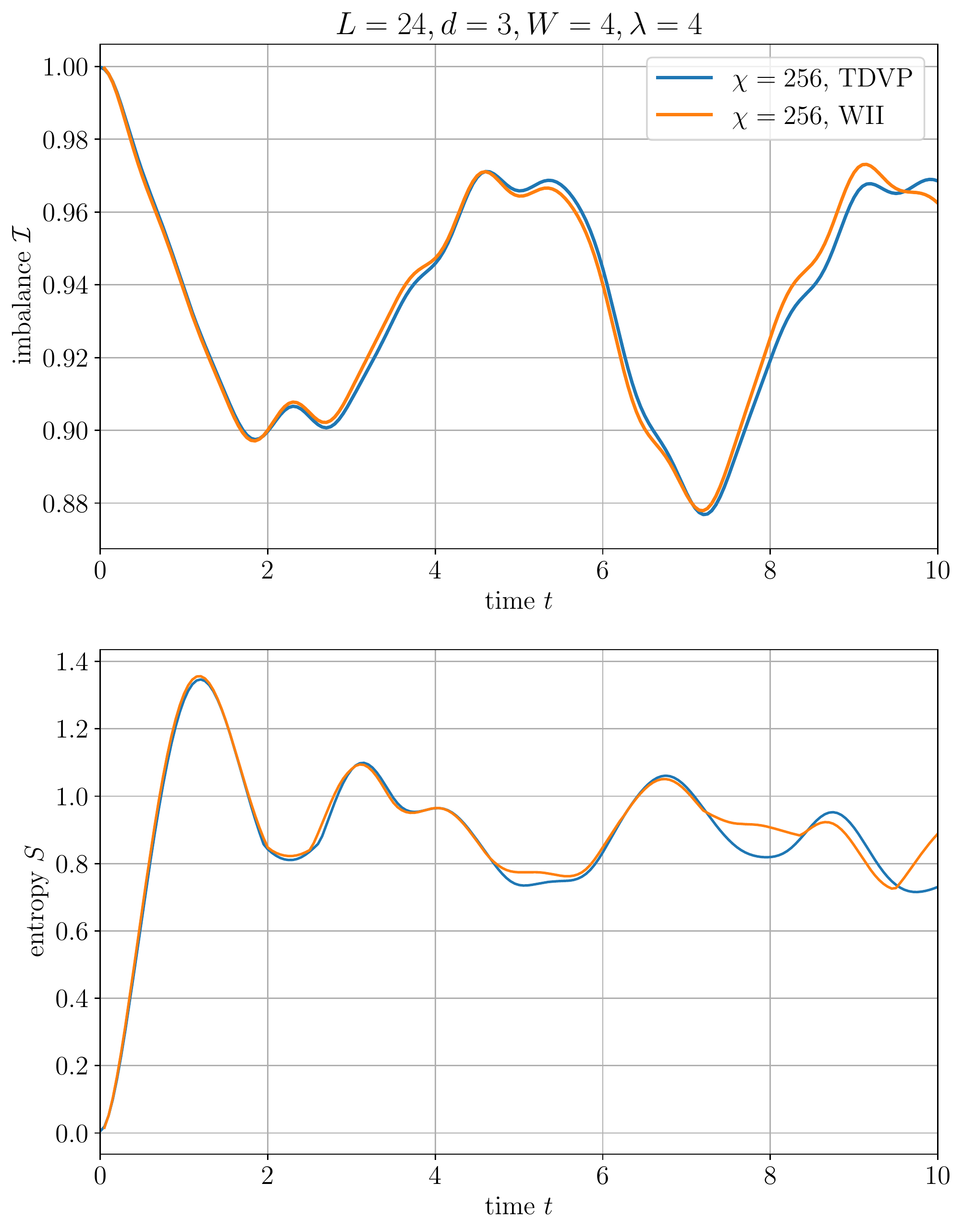}
    \caption{Comparison of the TDVP and WII algorithms over the time window $t \in [0, 10]$.}
    \label{fig:wiibench}
\end{figure}

\subsection{Comparison to the $W_\mathrm{II}$ method}

We now compare the results of the TDVP algorithm, used in the main text, to a different method also belonging to the class of MPS algorithms: the $W_\mathrm{II}$ method \cite{Zaletel2015a} (we again use the TeNPy library \cite{tenpy} to implement it). This method shares some features of the TDVP algorithm, such as the ability to handle long-range terms. The latter is an essential ingredient used for the mapping of the two-dimensional square lattice to the 1D structure of the MPS. A key difference between the methods is that the $W_\mathrm{II}$ method suffers from a truncation error at each time step, associated with truncated singular value decompositions. This type of error does not appear in single-site TDVP; however, a projection error, induced by the projector $\mathcal{P}_\mathrm{MPS}$ takes its place. Moreover, single-site TDVP is an implicit integration method; such methods tend to be suitable for oscillatory problems. The reader is referred to the review \cite{Paeckel2019a} for a detailed discussion of the differences between these two algorithms, and to the review \cite{Doggen2021a} for an in-depth discussion of the application of MPS-type algorithms to the MBL problem. In the particular case of our model, we find that the TDVP performs better in terms of computational time, primarily because a smaller time step of the integrator is required for the $W_\mathrm{II}$ method, $\delta t = 0.01$ as opposed to $\delta t = 0.05$ for the TDVP.

The result is shown in Fig.~\ref{fig:wiibench}. At short times, the two methods are in good agreement, but the agreement deteriorates over time; the maximum bond dimension $\chi = 256$ is reached at $t \approx 1$, and repeated truncation errors cause the result to diverge from the TDVP method.

\subsection{Exact results}

\begin{figure}[htb]
    \centering
    \includegraphics[width=\columnwidth]{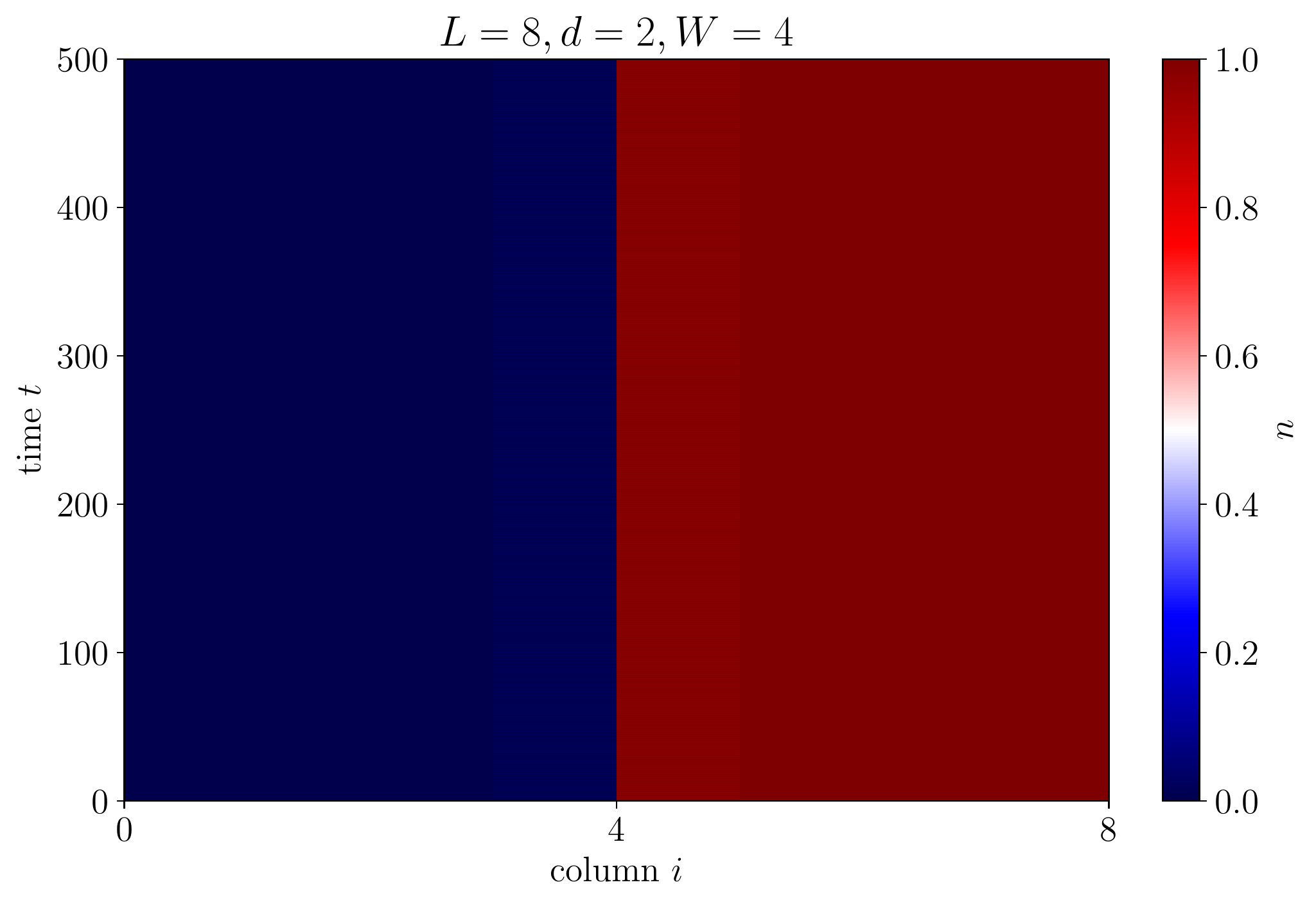}
    \caption{Numerically exact dynamics of the density, for $L = 8$, $d = 2$, $W = 4$, and $ \lambda = 4$.}
    \label{fig:exact}
\end{figure}

For sufficiently small system sizes, we can compare to numerically exact results, where there is no truncation of the Hilbert space through $\chi$. In Fig.~\ref{fig:exact}, we show such results, where we take $L = 8$, $d=2$, $W =4$, and $\lambda = 4$. The dynamics is almost entirely frozen, with the imbalance (not shown) saturating at $\mathcal{I} \approx 0.993$. This is because the orientation of the single domain wall is from unoccupied sites on the left side to occupied sites on the right side. As seen in the main text, it is the opposite orientation that allows for a greater degree of domain wall melting. The symmetry is broken through the sign of the hopping $J$.

\bibliography{ref}

\end{document}